\documentclass[debug,overfull]{epl}
\usepackage{epsfig}

\title{
Raman signatures of classical and quantum phases in coupled dots:
A theoretical prediction}
\shorttitle{Quantum phases in coupled dots}
\author{Massimo Rontani\thanks{E-mail: \email{rontani@unimo.it}}
\and Guido Goldoni \and
Franca Manghi \and Elisa Molinari\thanks{Group website:
\email{http://www.nanoscience.unimo.it}}}
\shortauthor{Massimo Rontani \etal}
\institute{
INFM and Dipartimento di Fisica,
Universit\`a degli Studi di Modena e Reggio Emilia -
via Campi 213/A, 41100 Modena, Italy
}
\begin{document}

\maketitle

%
\begin{abstract}
We study electron molecules in realistic vertically coupled quantum dots in a
strong magnetic field. Computing the
energy spectrum, pair correlation functions, and dynamical form
factor as a function of inter-dot coupling
via diagonalization of the many-body Hamiltonian, we
identify structural transitions between different phases, some of
which do not have a classical counterpart.
The calculated Raman cross section shows how such phases can
be experimentally singled out.
\end{abstract}



Electron systems form a Wigner crystal at sufficiently low density
or high magnetic field $B$\cite{WC}. Theoretical\cite{tanatar,lam}
and experimental\cite{WCexp,IP} studies suggest that lowering
dimensionality favors localization: in this perspective
interacting electrons
confined in a quantum dot (QD)\cite{review}, sometimes called
Wigner molecules\cite{WM}, are interesting in their own
right\cite{lit,filinov}, due to the interplay between the
electron-electron repulsion and the confining potential. This
leads to a complex zero temperature phase diagram\cite{bolton}, as
compared to the infinite layer case, as well as to complex melting
mechanisms\cite{filinov}.
The formation of coupled QDs (artificial molecules) introduces
qualitatively new physics\cite{leo}. New energy scales appear
--- inter-dot tunneling, inter- vs
intra-dot Coulomb correlation ---, whose balance controls the
phase diagram\cite{me}. Significantly, these parameters can be
tuned by inter-dot distance $d$ and/or electron
density, so that the nature of these few-particle systems and their
phases can be explored experimentally.

In this Letter we discuss quantum mechanical calculations of $N$
electrons in a coupled QD structure in the strong field
regime, where localization is ensured in the parent isolated
QDs. Monitoring the spatial correlation functions, we identify
different ground states depending on the inter-dot coupling. More
interestingly, we find that some of the phases do not have a classical
counterpart\cite{classic}, and are ascribed to a two-three-two
dimensional (2D-3D-2D) transition of the electronic system. Such
phases were not identified in previous studies of the coupled
layer system due to the neglect of tunneling and/or finite width
of the layers\cite{palacios,maksym,bart,lozovik2}, i.e., of the 3D
nature of the system, which turns out to be essential for their
formation. From the analysis of the dynamical form factor, we
associate to each phase peculiar collective charge excitations. By
explicit calculation of the optical spectra, we predict that Raman
selection rules and peak positions may clearly discriminate
between the different phases.

We consider two vertically coupled QDs, defined by
the potential $V(\vect{r})=m^*\omega_0^2\varrho^2/2+V(z)$
given by the sum of an in-plane term [$\vect{\varrho}\equiv (x,y)$,
$m^*$ effective mass, $\omega_0$ characteristic
frequency], and a profile $V(z)$ along the growth direction $z$,
which is a symmetrical square double quantum well (DQW). Each
well, of width $L_{\rm w}$ and potential height $V_0$, contains
one of the two QDs. They are separated by a barrier of width $d$.
The single-particle Hamiltonian in the symmetric gauge is
$H_0(\vect{r},s_z)=(-{\rm
i}\hbar\nabla+\left|
e\right|\vect{A}/c)^2/2m^*+V(\vect{r})+g^*\mu_BBs_z$, with $\vect{B}$ parallel
to the $z$-axis ($\vect{A}=\vect{B}\times\vect{\varrho}/2$, $\mu_B$ Bohr
magneton, $g^*$ effective giromagnetic factor, $s_z=\pm 1/2$
spin).
Since we are interested in the strong localization regime of small
filling factors $\nu$, we expect that spin texture does not alter the
essential physics. Therefore
we assume that electrons are spin polarized\cite{tejedor} and neglect
Landau level mixing\cite{review,WM,palacios}.
The eigenfunctions of $H_0$ are
$\psi_{m,g}(\vect{r})=\varphi_m(\vect{\varrho})\, \chi_i(z)$,
where $\varphi_m(\vect{\varrho})$ ($m=0,1,2,\ldots$) are
the Fock-Darwin orbitals of the first Landau band,
and $\chi_1$ and  $\chi_2$ the symmetrical (S, bonding)
and antisymmetrical (AS, antibonding) states of the DQW, respectively.
We neglect higher subbands since in real QDs the
confinement in the $z$ direction is stricter than in the $xy$
plane.
The index $g$ in $\psi_{m,g}(\vect{r})$ stands for the
parity under spatial inversion $\vect{r}\rightarrow -\vect{r}$:
even when $m+i$ is odd, odd otherwise. The many-body
Hamiltonian $\cal{H}$ [$\alpha\equiv (m,g)$] is:
\begin{displaymath}
{\cal{H}}=\sum_{\alpha}\varepsilon_{\alpha}c^{\dagger}_{\alpha}c_{\alpha}
+\frac{1}{2}\sum_{\alpha\beta\gamma\delta}
\Big<\alpha\beta\Big|\frac{e^2}{\kappa_r\left|\vect{r_1}-\vect{r_2}\right|}
\Big| \gamma\delta\Big>c^{\dagger}_{\alpha}c^{\dagger}_{\beta}
c_{\gamma}c_{\delta}.
\end{displaymath}
$c_{\alpha}$ destroys an electron occupying the orbital $\alpha$. Here
the single particle energy $\varepsilon_{mg}=\hbar\Omega(m+1)
-\hbar\omega_c\,m/2 + \epsilon_i -|g^*|\mu_BB/2$ is the sum of the
in-plane contribution and the energy of the
$i$-th DQW state $\epsilon_i$, $\Omega=(\omega_0^2 +\omega_c^2/4)^{1/2}$,
$\omega_c=|e|B/cm^*$ is the cyclotron frequency,
$\kappa_r$ is the dielectric constant\cite{review}.
$\cal{H}$ is represented in a basis of Slater determinants
spanned by filling with $N$ electrons the single-particle
states $\psi_{m,g}$; it is diagonalized on each Hilbert space sector
labeled by the total orbital angular momentum $M$ and parity.

Since the basis must be truncated, as in any Configuration
Interaction approach, the Fock space was built by filling up to 32
orbitals, chosen within the set $\{\psi_{m,g}\}_{m,g}$ to minimize
the energy. In order to improve the accuracy of results, the
choice of single-particle orbitals $\psi_{m,g}$ depended on the
$M$-sector: by trial-and-error, we found two optimized sets of
orbitals: for $M\ge 70$, we included S levels with $m=0,\ldots,24$
and AS levels with $m=0,\dots,6$, while for $M\le 69$ we included
S levels with $m=3,\dots,17$ and AS levels with $m=0,\dots,16$.
Subspaces obtained in this way (with maximum size $\approx 2 \cdot
10^4$) were diagonalized via Lanczos method. The two-body Coulomb
matrix elements of $\cal{H}$ were computed numerically.

We consider $B$ fields such that single-QD
correlation functions show strong localization\cite{parameters}.
We present results for $N=6$.
The system with $N<6$ exhibits the same physics.
We can identify three
regimes which, in general, correspond to different
electron arrangements: (I) At small $d$,
tunneling dominates and the system behaves as a unique coherent
system. (II) As $d$ is increased, all
energy scales become comparable. (III) When eventually tunneling is
suppressed, only the ratio between intra- and inter-dot interaction
is the relevant parameter for the now well separated QDs.

\begin{figure}
\centerline{\epsfig{file=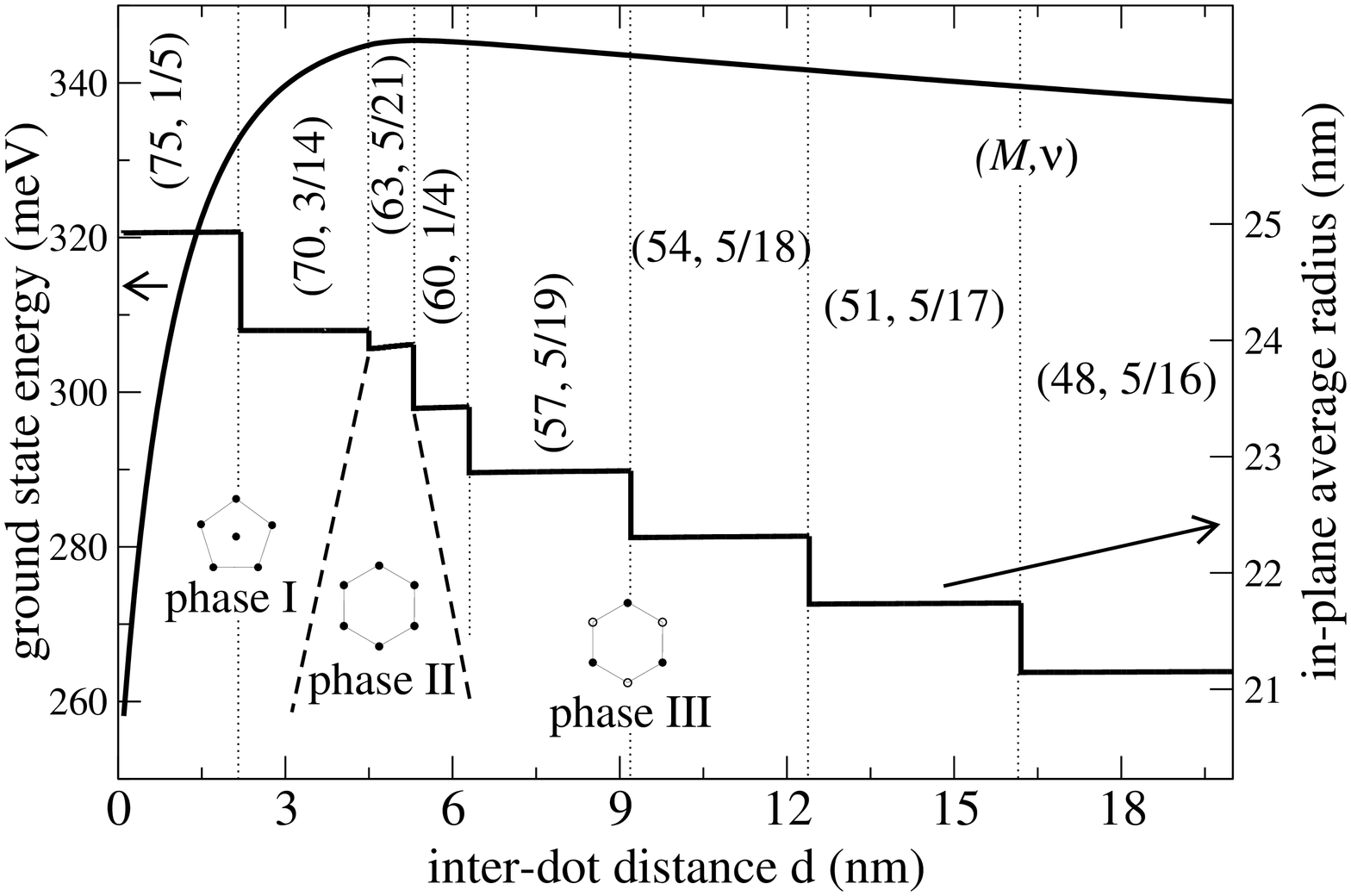,angle=90,width=17.6cm}}
\caption{
Ground-state energy (left axis) and
in-plane average radius $\left<\varrho\right>$ (right axis)
vs $d$ for $N=6$ at $B=25\,\mbox{T}$.
For large $d$-values $M=36$ (not shown), twice than in a single QD
with $N=3$. Insets show
the electron arrangements in the different phases.
}
\label{fig1}
\end{figure}

Figure \ref{fig1} shows the
calculated ground-state energy vs $d$. At small
distances the curve increases with $d$ (phase I), because the
kinetic energy exponentially grows due to the progressive
localization of the wavefunction into the dots: electrons occupy
only S orbitals, whose energies increase. Close to
$d=5$ nm (phase II), inter-dot tunneling is suppressed, so that
S and AS orbitals are almost degenerate: the
dominant energy contribution in this regime is the inter-dot
Coulomb interaction $\propto 1/d$ (phase III). As $d$ varies,
ground- and first-excited states, labelled by different $M$'s and
parities, cross each other at several critical values (thin
vertical lines).
The monotonous decrement of $M$\cite{palacios}
proceeds by steps of 5 and then 3 units, in
regions I and III, respectively;
as discussed later, these are ``magic numbers''\cite{WM}, reflected
in the spectrum of collective excitations.
Correspondingly, the in-plane
average radius $\left<\varrho\right>$ (right axis in
Fig.~\ref{fig1}) has a stair-like behavior and it
is almost constant at a given $M$.
$M$ measures the in-plane spatial
extension of the charge density: the higher $M$ the outer orbitals
occupied\cite{maksymprl}. We define a total filling factor $\nu$,
in analogy with double layers in the fractional quantum Hall
effect (FQHE), as $\nu=N(N-1)/2M$\cite{maksymprl}. Here
$\nu$ ranges from 1/5 at $d=0$ to 5/12 when the two dots are
isolated (then $\nu /2=5/24$ refers to a single
dot with $N=3$).

The flat steps of $\left<\varrho\right>$ in
Fig.~\ref{fig1} imply that these states are incompressible, in the
same sense as Laughlin's states of the FQHE\cite{bob}. Indeed,
varying $d$ acts like an external pressure applied in the $z$
direction, forcing the wavefunction to change: however, due to a
cusp-like structure of the energy spectrum\cite{maksymprl}, this
happens only in a discontinuous way. In the tunneling-dominated
regime, up to $d=4.5$ nm (phase I), increasing $d$ means enlarging
the volume of basically a unique QD, thus forcing a rearrangement
of the incompressible charge density at the $M$-transition
$75\rightarrow 70$. In the ``compressible'' region
II the penetration of the wavefunction into the inter-dot barrier
goes to zero: the effective volume of the charge density is thus
reduced, which explains the slight and
continuous increase of $\left<\varrho\right>$ with $d$, as shown
in Fig.~\ref{fig1}.
For $d> 5.3$ nm (phase III), the dots are well
separated, and $\nu$ increases from 1/4 to 5/12 when $d\rightarrow
\infty$, which is well known from double-layer
physics\cite{IP,guido}, due to decrease of inter-dot correlation
(stabilizing the Wigner molecule).

\begin{figure}
\centerline{\epsfig{file=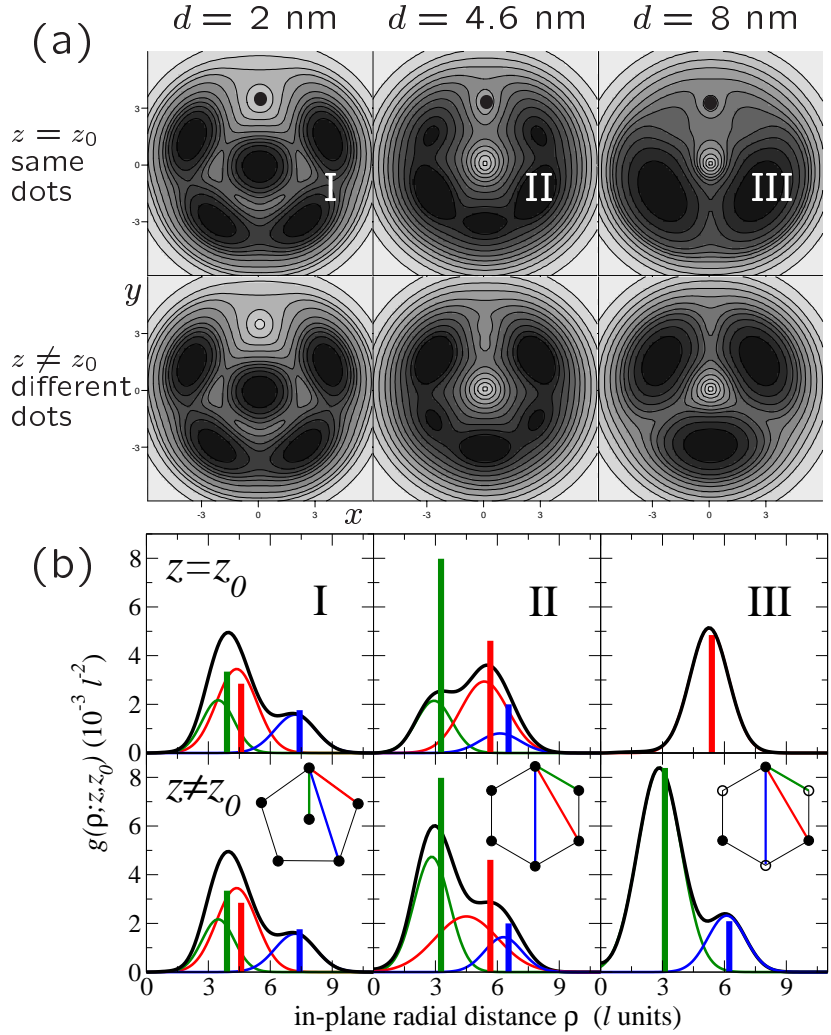,width=13.7cm}}
\caption{
(a) Contour plot of the pair correlation function
$P( \vect{\varrho} , z ; \vect{\varrho}_0 , z_0 )$
at various $d$-values.
An electron is fixed at $(\vect{\varrho}_0 , z_0 )$ (black bullet)
corresponding to a maximum of charge density. (b) Radial pair
correlation function $g(\varrho;z,z_0)$ vs $\varrho$. $g$ has been
renormalized in such a way that $\int\! {\rm d}\vect{\varrho}\,
g(\varrho;z,z_0) = 1$. Lengths are in units of $\ell=7.20$ nm
[$\ell=(\hbar/m^*\Omega)^{1/2}$ characteristic in-plane radius].
Insets schematically show the spatial arrangement of electrons projected
on the $xy$ plane.
}
\label{fig2}
\end{figure}

To analyze the internal structure of the molecule, we compute the
pair correlation function $P( \vect{\varrho} , z ; \vect{\varrho}_0
, z_0 ) =\sum_{i\neq j}\left<\delta
(\vect{\varrho}-\vect{\varrho_i}) \delta ( z - z_i ) \delta
(\vect{\varrho_0}-\vect{\varrho_j}) \delta (z_0 - z_j)
\right>/N(N-1)$ (the average is on the ground state). Figure
\ref{fig2}(a) shows a contour plot of $P( \vect{\varrho} , z ;
\vect{\varrho}_0 , z_0 )$ at various values of $d$, one per
column. An electron is fixed (black bullet) at the position
$(\vect{\varrho}_0,z_0)$ in one dot, at the maximum of the
charge density: the contour plots of the top (bottom) row, with
$z=z_0=$ dot 1 ($z=$ dot 2, $z_0=$ dot 1) fixed, represent the
conditional probability of measuring an electron in the $xy$ plane
of dot 1 (2), given a first electron is fixed in dot 1. We find
that the three phases we discussed above clearly show
characteristic structural configurations (see also insets in
Fig.~1): (I) At small $d$ (left column), the whole system is
coherent. The electrons, delocalized over the dots, arrange at the
vertices and the center of a regular pentagon. (II) At a critical
value of $d=4.6$ nm (center) an abrupt transition takes
place. In the new arrangement the electrons are at the vertices of
a regular hexagon. Contrary to phase I, the peaks in the upper and
lower dots have different heights. (III) When $d$ is further
increased, the structure continuously evolves into two isolated
dots (right), coupled only via Coulomb interaction. Three
electrons in each dot sit at the vertices of two equilateral
triangles rotated by 60 degrees.

Next we compare our system with its classical
counterpart\cite{classic}. To this aim, we calculated the
(circularly symmetric) radial pair correlation function
$g(\varrho; z,z_0)= \int\! {\rm d}\vect{\varrho_0}\, P(
\vect{\varrho_0}+ \vect{\varrho},z; \vect{\varrho_0},z_0)$ vs
$\varrho$, which gives the probability of finding an electron at a
distance $\varrho$ from another one fixed on the same or opposite
dot. For a classical system at zero temperature, $g$ consists of a
set of $\delta$-function peaks. Figure \ref{fig2}(b) shows the
calculated $g(\varrho; z,z_0)$ for the same three configurations
of \ref{fig2}(a), together with the histograms representing the
equilibrium configuration of the corresponding classical
system\cite{classic,gr2}. When we fit $g$ to a set of gaussians,
also shown in Fig.~\ref{fig2}(b), we see that the agreement
between ``classical'' and ``quantum'' cases is remarkable in phase
III and, to a minor extent, in phase I,  both from the point of
view of absolute positions and relative intensities of the peaks.
In contrast, in phase II the peaks of the quantum system resemble
the classical ones just in position, while the intensities are
definitely different; calculations show, moreover, that the
relative intensities are strongly $d$-dependent. Classically, only
phases I and III are ground-state configurations\cite{classic},
while the hexagonal structure II is a metastable
state\cite{bolton}. Quantum fluctuations thus force a novel phase
to appear.

We interpret the sequence I $\rightarrow$ II $\rightarrow$ III as
a 2D $\rightarrow$ 3D $\rightarrow$ 2D transition. In phase III
the motion is quasi-2D, with electrons occupying degenerate S and
AS states to form a staggered configuration: the equilibrium
structure corresponds to the classical one, with inter-dot
tunneling absent. In phase II, instead, the motion acquires an
effective $z$-component: as $d$ is decreased, electrons keep their
electrostatic repulsion energy low at the expense of occupying AS
states separated from S orbitals by an increasing energy gap.
There is no classical analogue to this phase. When the gap becomes
too large, the electrons abruptly arrange into I, all frozen in S
orbitals. The motion now is that of a coherent quasi-2D QD,
comparable to a classical single QD. To sum up, only in phase II
electrons move in a truly 3D system: as shown in
Fig.~\ref{fig2}(a), the angular modulation of $P$ is much weaker
than in I and III, suggesting that the crystallization regime has
not been reached yet\cite{tanatar,lam}.

There is another, more subtle, discrepancy between quantum and
classical cases, due to the interplay of tunneling and particle
indistinguishability. In phase I classical particles belong either
to one dot or to the other, three each for $N=6$, and they can
arrange in very asymmetric configurations in each dot, 
like in Fig.~1 of Ref.~\cite{classic}, provided
that overall they form the centered pentagon which is favored by
Coulomb interaction, the only energy scale in this case. In the
quantum case, on the contrary, particles are equally delocalized
on both dots. It simply makes no sense in this phase to assign
electrons to a specific dot, since at small $d$ they can easily
penetrate the inter-dot barrier. Therefore, tunneling changes the
physics of the artificial molecule and actually drives phase I.
Indeed, if we suppress tunneling (e.g., via a sufficiently high
barrier), phase III is the only one present at any value of $d$,
while phase I never shows up, contrary to the classical
prediction\cite{classic}.

\begin{figure}
\centerline{\epsfig{file=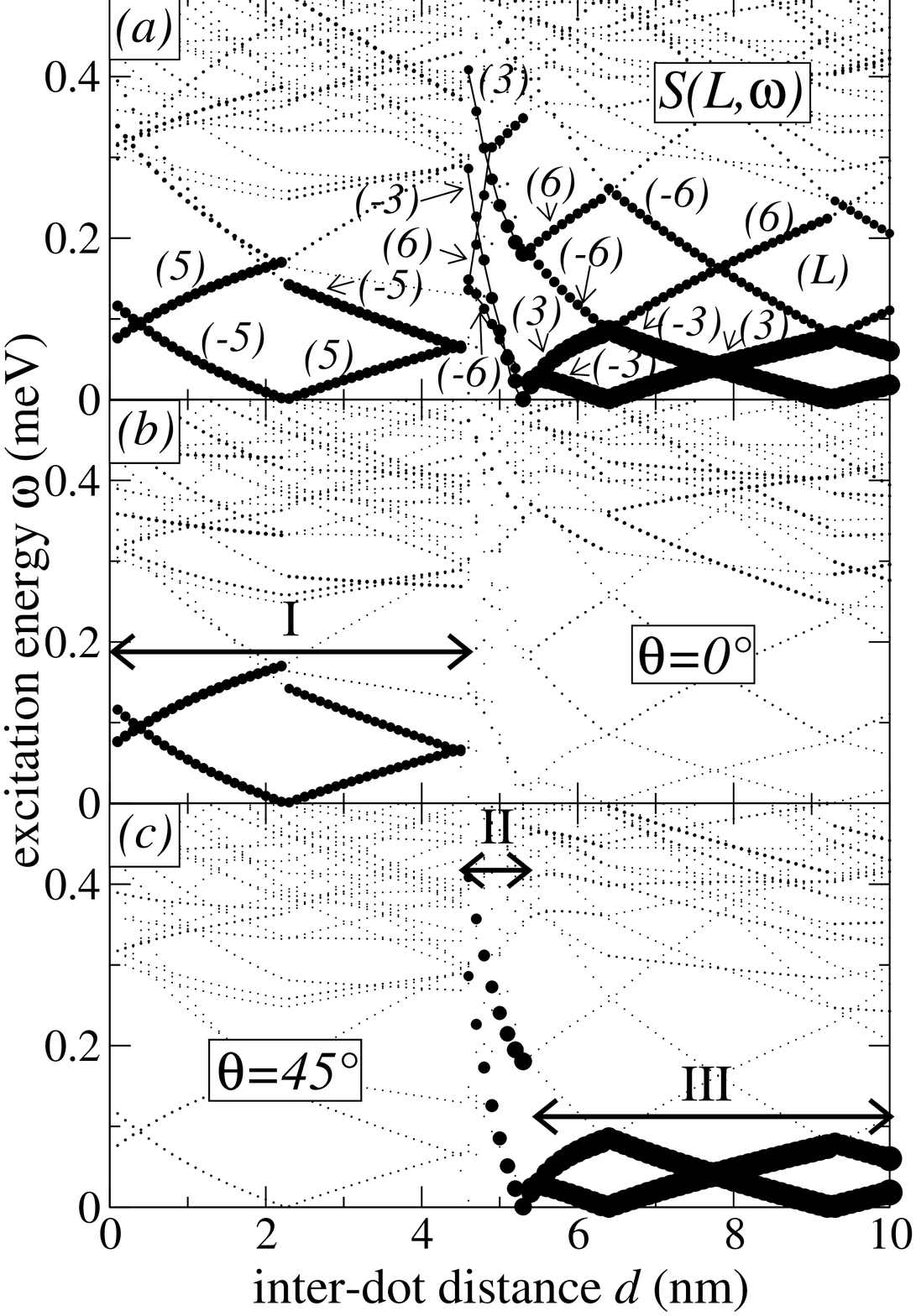,width=13.6cm}}
\caption{
(a) Dynamical form factor $S(L,\omega)$ in the $d- \omega$ plane,
for different angular momentum components $L$, indicated
in parentheses. The radius of each
point is proportional to its intensity.
(b) Analogous plot
for the Raman scattering cross section, at an angle $\theta=0^{\circ}$
between the $xy$ plane and the transferred momentum $\vect{q}$
($q=2\cdot 10^{6}$ cm$^{-1}$).
(c) Same as (b), with $\theta=45^{\circ}$.
}
\label{fig3}
\end{figure}

We analyze the spectrum of neutral collective elementary
excitations computing the dynamical form factor
$S(L,\omega)=\sum_n|\langle n|\rho^{\dagger}_L|0\rangle |^2
\,\delta\!\left(\omega-\omega_n+\omega_0\right)$, with
$\left|n\right>$ ($\left|0\right>$) many-body $n$-th excited
(ground) state with energy $\omega_n$ ($\omega_0$), $\omega$
excitation energy, $\rho^{\dagger}_L=\sum_{m}\sum_{g,g^{\prime}}
c^{\dagger}_{L+m,g}c_{m,g^{\prime}}$ $L$-th angular component of
density fluctuation. $S(L,\omega)$ assigns a weight to a charge
density wave of angular momentum $L$ and energy $\omega$ in the
excitation spectrum. Figure \ref{fig3}(a) shows $S(L,\omega)$ at
low energies in the $d-\omega$ space; the intensity is
proportional to the radius of circles. Only few branches are
dominant in each phase, with a characteristic value of $|L|$: 5 in
I, 6 or 3 in II and III, respectively.
The occurrence of very few excitations with large values of
$S(L,\omega)$\cite{valueS} is remarkable:
since the ground state is formed by a linear combination of
thousands of Slater determinants with non-negligible weight,
high values of the form factor are due to a constructive interference
effect. Indeed, the geometry
of the Wigner molecule selects the allowed --magic-- values of $M$
and parity, and thus $L=M_2-M_1$
corresponding to the coupling between two magic states $M_1$,
$M_2$. Consider, e.g., the wavefunction $\Psi$ of phase II with
electrons at the vertices of a regular hexagon. A cyclic
coordinate permutation is a $\pi/3$ rotation such that
$\Psi\rightarrow \exp{(\pi{\rm i}M/3)}\Psi$. In this case
$\Psi$ changes sign\cite{magic2}, and therefore $M=6p+3$
($p$ integer)
and odd parity. Our findings show that an excited magic state
can be created adding
$L$ quanta of angular momentum to the ground state via the density
fluctuation operator $\rho^{\dagger}_L$. The allowed values of $L$
characterize different phases. The diamond shape of different
$|L|$-branches in Fig.~\ref{fig3}(a) originates from level
crossing between magic states: $\omega\rightarrow 0$ at every
$M$-transition in the ground state.

Finally, we compute the ``electronic'' part of Raman scattering cross
section\cite{hawrylak}.
This is strictly related to $S(L,\omega)$, and is proportional to
$\sum_n\left|M_{n0}\right|^2\delta\left(\omega-\omega_n+\omega_0\right)$,
with $M_{n0}=\sum_{\alpha,\beta} \left<\alpha|{\rm e}^{{\rm
i}\vect{q}
\cdot\vect{r}}|\beta\right>\left<n|c^{\dagger}_{\alpha}c_{\beta}
|0\right>$, and $\vect{q}$ wave vector transferred in the inelastic
photon scattering\cite{ramanth}. The correlated excitation
spectrum of single QDs has been already experimentally
probed\cite{ramanexp}, in the regime $q\ell\approx 1$. Thus,
having fixed a suitable value of $q$ in a backscattering geometry,
in Figs \ref{fig3}(b) and (c) we plot the cross sections at
different angles $\theta$ between $\vect{q}$ and the $xy$ plane
($\theta=0^{\circ}$ and $45^{\circ}$, respectively). The spectra
resemble the form factor $S(L,\omega)$, but
intensities are differently modulated. For in-plane scattering
[Fig.~\ref{fig3}(b), $\theta=0^{\circ}$], only the branches with
$|L|=5$ characterizing phase I survive, while all other signals
are suppressed.
The background small dots appearing in Fig.~\ref{fig3}(b-c)
are smaller than 9 \% of the maximum value.
As $\theta$ increases [Fig.~\ref{fig3}(c),
$\theta=45^{\circ}$], the branches of phase I continuously loose
intensity as those of II and III acquire weight, until the
latter show a very strong signal at $\theta=45^{\circ}$, while the
former are suppressed.
In I, indeed, all electrons occupy S orbitals, hence
the $z$-component $\int\!{\rm d}z \,\chi^*_i(z)\, {\rm e}^{{\rm
i}q_z z}\chi_j(z)$ of matrix element $\left<\alpha|{\rm e}^{{\rm
i}\vect{q} \cdot\vect{r}}|\beta\right>$ diminishes as long as
$\theta$ (and $q_z$) increases; the opposite holds true for II and
III, since S and AS orbitals can mix. This means
that momentum can be transferred in the $z$-direction only in
II and III, contrary to I, where the system is a unique
quasi-2D dot. Signals with $|L|=6$ are almost invisible, but still
phase II can be resolved from III in virtue of the characteristic
energy and slope of visible branches. This comes from the way S and
AS orbitals are differently filled in II and III,
respectively, and hence from the different contribution of
tunneling to the total energy. When $\theta$ is further increased,
the selection rule $L=0$ becomes effective, and eventually at
$\theta=90^{\circ}$ low-energy signals are suppressed.

To summarize, we have predicted a 2D-3D-2D transition of
interacting electrons in a double QD, accompanied by the
appearance of a novel liquid-like quantum phase, in addition to
classical configurations. The $\theta$-dependent modulation of the
Raman spectrum should allow to experimentally discriminate between
different phases. \acknowledgments Work supported by INFM through
PRA SSQI. We thank for useful discussions V.~Pellegrini and
A.~Pinczuk.


%
\clearpage
%
%

\begin{thebibliography}{0}
\bibitem{WC}
\Name{Wigner E.}
\REVIEW{Phys. Rev.}{46}{1934}{1002};
\Name{Ishiara A.}
\REVIEW{Solid State Phys.}{42}{1989}{271};
\Name{C\^ot\'e R.}
\Book{Microscopic Theory of Semiconductors}
\Editor{S. W. Koch}
\Publ{World Scientific, London}
\Year{1995}
\Page{315}.
\bibitem{tanatar}
\Name{Ceperley D. M. \and Alder B. J.}
\REVIEW{Phys. Rev. Lett.}{45}{1980}{566};
\Name{Tanatar B. \and Ceperley D. M.}
\REVIEW{Phys. Rev. B}{39}{1989}{5005}.
\bibitem{lam}
\Name{Lam P. K. \and Girvin S. M.}
\REVIEW{Phys. Rev. B}{30}{1984}{473}.
\bibitem{WCexp}
\Name{Grimes C. C. \and Adams G.}
\REVIEW{Phys. Rev. Lett.}{42}{1979}{795}.
\bibitem{IP}
\Name{Manoharan H. C., Suen Y. W., Santos M. B. \and
 Shayegan M.}
\REVIEW{Phys. Rev. Lett.}{77}{1996}{1813}.
\bibitem{review}
\Name{Jacak L., Hawrylak P., \and W\'ojs A.}
\Book{Quantum Dots}
\Publ{Springer, Berlin}
\Year{1998}.
\bibitem{WM}
For a review see
\Name{Maksym P. A., Imamura H., Mallon G. P. \and Aoki H.}
\REVIEW{J. of Phys.: Cond. Mat.}{12}{2000}{R299}.
\bibitem{lit}
\Name{Bryant G. W.}
\REVIEW{Phys. Rev. Lett.}{59}{1987}{1140};
\Name{Yannouleas C. \and Landman U.}
\SAME{82}{1999}{5325};
\SAME{85}{2000}{2220(E)};
\Name{Egger R., H\"ausler W., Mak C. H. \and Grabert H.}
\SAME{82}{1999}{3320};
\SAME{83}{1999}{462(E)};
\Name{Reimann S. M., Koskinen M. \and Manninen M.}
\REVIEW{Phys. Rev. B}{62}{2000}{8108};
\Name{M\"uller H.-M. \and Koonin S. E.}
\SAME{54}{1996}{14532}.
\bibitem{filinov}
\Name{Filinov A. V., Bonitz M. \and Lozovik Yu. E.}
\REVIEW{Phys. Rev. Lett.}{86}{2001}{3851}.
\bibitem{bolton}
\Name{Bolton F. \and R\"ossler U.}
\REVIEW{Superlatt. and Microstruct.}{13}{1993}{139};
\Name{Bedanov V. M. \and Peeters F. M.}
\REVIEW{Phys. Rev. B}{49}{1994}{2667}.
\bibitem{leo}
\Name{Kouwenhoven L.}
\REVIEW{Science}{268}{1995}{1440}.
\bibitem{me}
\Name{Rontani M., Troiani F., Hohenester U. \and Molinari E.}
\REVIEW{Solid State Commun.}{119}{2001}{309};
\Name{Rontani M., Rossi F., Manghi F. \and Molinari E.}
\SAME{112}{1999}{151}.
\bibitem{classic}
\Name{Partoens B., Schweigert V. A. \and Peeters F. M.}
\REVIEW{Phys. Rev. Lett.}{79}{1997}{3990}.
\bibitem{palacios}
\Name{Palacios J. J. \and Hawrylak P.}
\REVIEW{Phys. Rev. B}{51}{1995}{1769}.
\bibitem{maksym}
\Name{Imamura H., Maksym P. A. \and Aoki H.}
\REVIEW{Phys. Rev. B}{53}{1996}{12613};
\SAME{59}{1999}{5817}.
Authors identify phases I and III for $N\le 4$.
\bibitem{bart}
\Name{Partoens B., Matulis A. \and Peeters F. M.}
\REVIEW{Phys. Rev. B}{59}{1999}{1617}.
\bibitem{lozovik2}
\Name{Filinov A. V., Bonitz M. \and Lozovik Yu. E.}
\REVIEW{Contrib. Plasma Phys.}{41}{2001}{1/2}.
\bibitem{tejedor}
At larger $\nu$ spin plays an important role; for example,
in the regime close to total filling factor two, it may give rise
to a non trivial phase diagram
[\Name{Mart\'in-Moreno L., Brey L. \and Tejedor C.}
\REVIEW{Phys. Rev. B}{62}{2000}{R10633}]. 
\bibitem{parameters}
We assumed $B=25$ T, and $m^*=0.067m_e$, $\kappa_r=12.4$,
$g^*=-0.44$, $\hbar\omega_0= 3.70\,\mbox{meV}$, $L_{\rm w}= 12
\,\mbox{nm}$, $V_0= 250\,\mbox{meV}$, as typical values of
realistic devices
[Cf. \Name{Pi M., Emperador A., Barranco M., Garcias F., Muraki K.,
Tarucha S. \and Austing D. G.}
\REVIEW{Phys. Rev. Lett.}{87}{2001}{66801}].
\bibitem{maksymprl}
\Name{Maksym P. A. \and Chakraborty T.}
\REVIEW{Phys. Rev. Lett.}{65}{1990}{108}.
\bibitem{bob}
\Name{Laughlin R. B.}
\REVIEW{Phys. Rev. B}{27}{1983}{3383}.
\bibitem{guido}
\Name{\'Swierkowski L., Neilson D. \and ~Szyma\'nski J.}
\REVIEW{Phys. Rev. Lett.}{67}{1991}{240};
\Name{Goldoni G. \and Peeters F. M.}
\REVIEW{Europhys. Lett.}{37}{1997}{293}.
\bibitem{gr2}
For the classical histograms, we used data of a single QD
[\Name{Date G., Murthy M. V. N. \and Vathsan R.}
\REVIEW{J. of Phys.: Cond. Mat.}{10}{1998}{5876}]:
II differs from III in the way electrons are placed on a particular dot.
We normalized $\delta$-functions so that
the two highest classical and quantum peaks in III
have the same height.
\bibitem{valueS}
The value of dots not labeled by $(L)$ in Fig.~\ref{fig3}(a) is smaller than
16 \% of the maximum value.
\bibitem{magic2}
\Name{Ruan W. Y., Liu Y. Y., Bao C. G. \and Zhang Z. Q.}
\REVIEW{Phys. Rev. B}{51}{1995}{7942}.
\bibitem{hawrylak}
\Name{Hawrylak P.}
\REVIEW{Solid State Commun.}{93}{1995}{915}.
\bibitem{ramanth}
\Name{Blum F. A.}
\REVIEW{Phys. Rev. B}{1}{1970}{1125}.
\bibitem{ramanexp}
\Name{Lockwood D. J., Hawrylak P., Wang P. D., Sotomayor Torres C. M.,
Pinczuk A. \and Dennis B. S.}
\REVIEW{Phys. Rev. Lett.}{77}{1996}{354};
\Name{Sch\"uller C., Keller K., Biese G., Ulrichs E., Rolf L., Steinebach C.,
Heitmann D. \and Eberl K.}
\SAME{80}{1998}{2673}.

\end{thebibliography}
\end{document}